\documentclass[aip,apl,twocolumn,reprint]{revtex4-1}
\usepackage{graphicx}
\usepackage{dcolumn}
\usepackage{bm}
\usepackage{color}
\usepackage{tabularx}
\usepackage{array}
\usepackage{amsmath}
\usepackage{stmaryrd}  

\begin{document}

\title{Hybrid magnonics: physics, circuits and applications for coherent information processing}

\author{Yi Li}
\affiliation{Materials Science Division, Argonne National Laboratory, Argonne, IL 60439, USA}

\author{Wei Zhang}
\affiliation{Department of Physics, Oakland University, Rochester, MI 48309, USA}
\affiliation{Materials Science Division, Argonne National Laboratory, Argonne, IL 60439, USA}

\author{Vasyl Tyberkevych}
\affiliation{Department of Physics, Oakland University, Rochester, MI 48309, USA}

\author{Wai-Kwong Kwok}
\affiliation{Materials Science Division, Argonne National Laboratory, Argonne, IL 60439, USA}

\author{Axel Hoffmann}
\email{axelh@illinois.edu}
\affiliation{Department of Materials Science and Engineering, University of Illinois at Urbana-Champaign Urbana, IL 61801}

\author{Valentine Novosad}
\email{novosad@anl.gov}
\affiliation{Materials Science Division, Argonne National Laboratory, Argonne, IL 60439, USA}

\date{\today}

\begin{abstract}

Hybrid dynamic systems have recently gained interests with respect to both fundamental physics and device applications, particularly with their potential for coherent information processing. In this perspective, we will focus on the recent rapid developments of magnon-based hybrid systems, which seek to combine magnonic excitations with diverse excitations for transformative applications in devices, circuits and information processing. Key to their promising potentials is that magnons are highly tunable excitations and can be easily engineered to couple with various dynamic media and platforms. The capability of reaching strong coupling with many different excitations has positioned magnons well for studying solid-state coherent dynamics and exploiting unique functionality. In addition, with their gigahertz frequency bandwidth and the ease of fabrication and miniaturization, magnonic devices and systems can be conveniently integrated into microwave circuits for mimicking a broad range of device concepts that have been applied in microwave electronics, photonics and quantum information. We will discuss a few potential directions for advancing magnon hybrid systems, including on-chip geometry, novel coherent magnonic functionality, and coherent transduction between different platforms. As future outlook, we will discuss the opportunities and challenges of magnonic hybrid systems for their applications in quantum information and magnonic logic.

\end{abstract}

\maketitle

\section{Introduction}

Hybrid dynamic systems have recently attracted great attentions due to their applications in quantum information, communications and sensing \cite{KurizkiPNAS2015}. Thus hybrid systems provide a new paradigm for combining platforms and devices that can perform different tasks such as storing, processing and transmitting coherent states, particularly in quantum information. In addition, coherent interactions that are based on those bosonic excitations are not restricted to the quantum limit, but can be also observed in the classical regime because phase coherence can be maintained for all the excitation quanta at the lowest energy states. This significantly facilitates the research on the fundamental physics of strong coupling between different excitations as well as coherent manipulation and engineering of hybrid systems.

One important feature in hybrid dynamic systems is that the coherence of information being carried in dynamic excitations can be maintained while being transduced from one module to another. Coherence, which is defined as the preservation of phase in excitations between different modules, is limited by the decoherence rate $\kappa$, which can be determined by the frequency linewidth. Conversely, the transduction rate is determined by the coupling strength $g$ in the frequency domain. Therefore, in order to maintain coherence during transduction, $g$ needs to be larger than $\kappa$. For two coupled systems with decoherence rate $\kappa_1$ and $\kappa_2$, it is convenient to define the cooperativity of the system: $C=g^2/\kappa_1\kappa_2$. $C>1$ and $g>\kappa_1, \kappa_2$ defines the strong coupling regime in order to conduct any practical coherent information operation.

The introduction of magnons in hybrid systems started from the exploration of spin ensembles coupled to microwave photons \cite{KuboPRL2010,SchusterPRL2010}, where the spin-photon coupling strength can be enhanced by increasing the total number of spins $N$ proportional to $\sqrt{N}$ \cite{ImamogluPRL2009,WesenbergPRL2009}. Subsequently it was quickly found that due to their orders of magnitude higher spin densities, magnetic materials and their collective spin excitations, or magnons, can provide much stronger coupling \cite{FlattePRL2010,BauerPRB2015}, from a few megahertz in spin ensembles\cite{KuboPRL2010,SchusterPRL2010,ProbstPRL2013} to a few hundred megahertz in yttrium iron garnet (YIG) crystals \cite{HueblPRL2013,TabuchiPRL2014,ZhangPRL2014,GoryachevPRApplied2014,BhoiJAP2014}, along with cooperativity up to $10^3\sim 10^4$. This has quickly triggered broad attention and activities on new hybrid platforms with different microwave cavity designs, different magnetic systems and improved coupling \cite{BaiPRL2015,ZhangnpjQuantumInfo2015,LambertJAP2015,MaierFlaigPRB2016,KostylevAPL2016,MorrisSREP2017,BhoiSREP2017,BoventerPRB2018,AbdurakhimovPRB2019,McKenziePRB2019,LiPRL2019_magnon,HouPRL2019,WolzCommPhys2020}. Thus a research direction called cavity magnonics (or cavity spintronics, spin cavitronics) has been created \cite{HuPhysCanada2016,BhoiSSP2019,WangJAP2020}, which focuses on the new properties, functionality and engineering of the coherent interactions between magnons and microwave photons.

\begin{figure*}[htb]
 \centering
 \includegraphics[width=4.5 in]{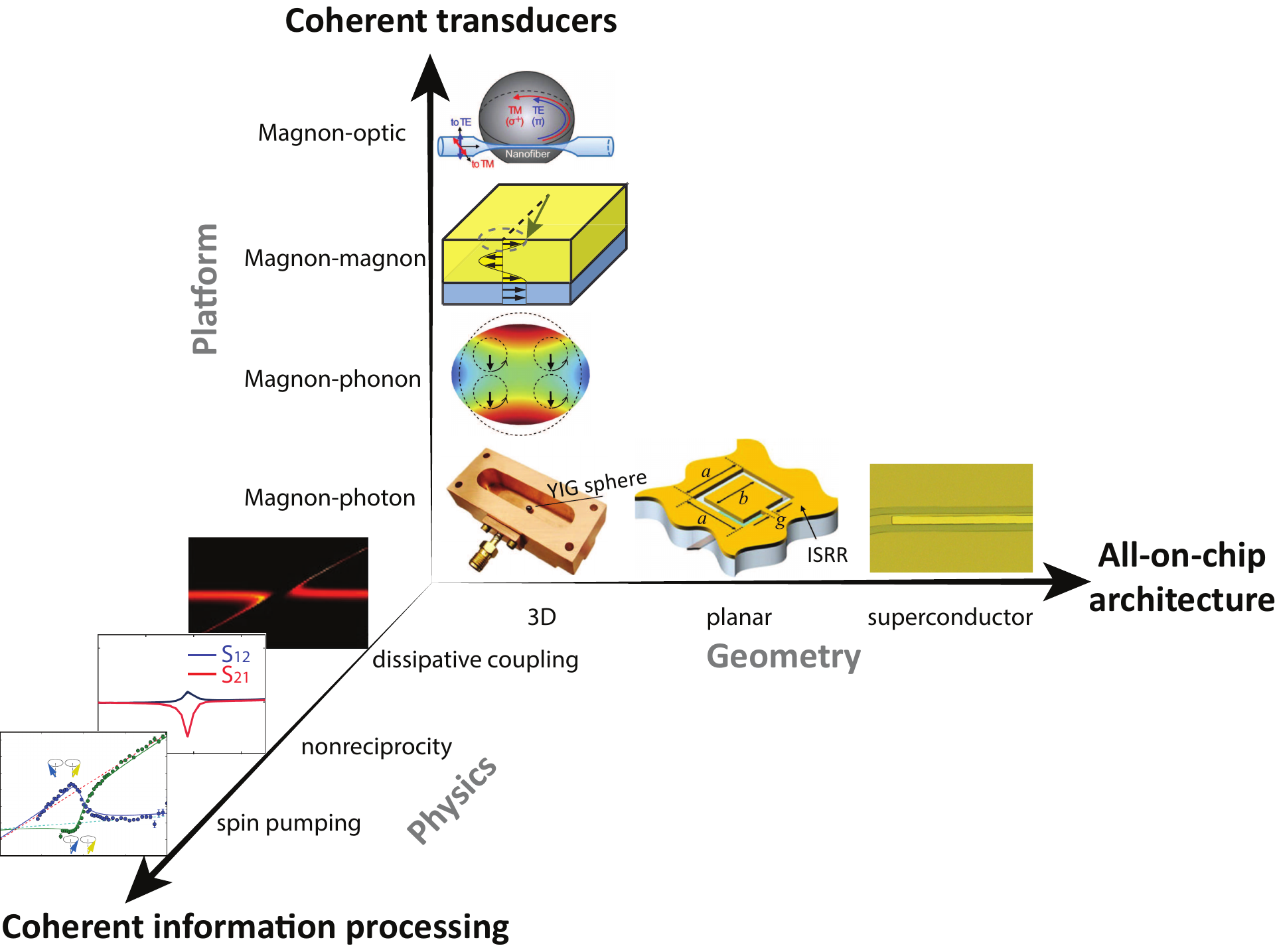}
 \caption{Growing directions and goals for magnon hybrid systems. The explorations on geometry, physics and platform lead to on-chip architecture, coherent information processing and coherent transduction, respectively.}
 \label{fig_overall}
\end{figure*}

A groundbreaking achievement came from Tabuchi, \textit{et al.} \cite{TabuchiScience2015}, who demonstrated coherent coupling between a single magnon and a superconducting qubit mediated by a microwave cavity. The same group has subsequently demonstrated detection of a single magnon by continuous-wave \cite{LachanceScienceAdvan2017} and single-shot (pulse) \cite{LachanceQuirionScience2020} inputs. These achievements open the possibility of conducting quantum information operations with magnons, leading to a new research field named quantum magnonics \cite{LachanceQuirionAPEx2019}. In addition, the capability of coupling microwave excitations with optical photons via magneto-optic interactions \cite{HaighPRA2015,OsadaPRL2016,ZhangPRL2016,HaighPRL2016} has generated interests of using magnons as quantum transducers between microwave quantum information and quantum optics \cite{LachanceQuirionAPEx2019,KusminskiyarXiv2019}. The potential of all-on-chip integration of magnon-photon hybrid systems, particularly with high-quality-factor superconducting resonators \cite{LiPRL2019_magnon,HouPRL2019}, can further boost the applications of magnon hybrid systems in quantum information.

Besides microwave photons, magnon hybrid systems have also been extended to other excitations, such as phonons, optical photons and magnons themselves, owing to the abundant coupling mechanisms associated with magnetic materials such as magneto-elastic, magneto-optic, dipolar and exchange couplings. The coherent interactions of different dynamic systems have inspired interdisciplinary explorations and provided new ideas for multi-stage coherent transduction of magnonic systems \cite{ZhangNComm2015,LambertPRA2016,ZhangScienceAdv2016,BaiPRL2017,LiJiePRL2018,AnPRB2020}.

In this perspective, we will discuss several emerging directions for the development of magnon hybrid system, which are summarized in Fig. \ref{fig_overall}. First, the integration of the magnon and microwave resonator systems, with geometry ranging from 3D to planar, provide new opportunity for device miniaturization. Second, the straightforward engineering of the magnon hybrid systems leads to new physics and properties which can be applied to new device concepts and functionality in coherent information processing. Third, numerous magnon-based hybrid systems have introduced new platforms for remote information transduction with magnons. Lastly, we will discuss how the discoveries of magnon hybrid systems can be applied in two particular areas, i.e., quantum information and magnonic logic.

\section{Magnon-photon coupling}

The interaction between magnons and microwave photons has been used for decades for measuring magnetization dynamics \cite{GriffithsNature1946,YagerPhysRev1947,KittelPhysRev1948}. The microwave excitations of magnons are typically achieved by placing magnetic samples into a microwave waveguide or cavity, and the ferromagnetic resonance signals can be measured from the transmission or reflection of microwave photons. The resonance signals describe the evolution of the magnetic susceptibility $\chi_m$, as:
\begin{equation}\label{eq01}
\chi_m = {\omega_M/2 \over (\omega_m-\omega) - i\alpha\omega_m}
\end{equation}
where $\omega_m$ is the magnon resonance frequency determined by the magnetic field, $\omega$ is the microwave frequency, $\alpha$ is the Gilbert damping and $\omega_M=\gamma \mu_0M_s$ with $\gamma$ the gyromagnetic ratio and $M_s$ the saturation magnetization. The factor $1/2$ comes from the linearly polarized microwave field coupled with circularly polarized magnetization precession. The demagnetizing field is ignored for simplicity.

Inside a microwave cavity, the magnetic sample becomes part of the cavity and contributes as an additional inductance. If we consider the microwave cavity as an \textit{LCR} resonator, the magnetic sample increases the total inductance as $L(\chi_m)=L_0(1+\chi_mV_m/V_c)$, where $L_0$ is the intrinsic inductance of the cavity, $V_c$ is the effective volume of the cavity, and $V_m$ is the volume of the magnetic system. The reflection of the microwave cavity can be described as
\begin{equation}\label{eq02}
  {P_{out} \over P_{in}} = {\kappa_R \over i(\omega_p-\omega)+\kappa_p}
\end{equation}
where $\omega_p=1/\sqrt{L(\chi_m)C}$ is the cavity resonant frequency, $\kappa_p=1/(2RC)$ is the damping rate of the \textit{LCR} resonator and $\kappa_R$ is the coupling rate to the external port. Eqs. (\ref{eq01}) and (\ref{eq02}) have a similar form because they describe the resonance of magnons and photons. In the limit of linear magnetic coupling, $V_m\ll V_c$, by expanding the $\chi_m$ term in Eq. (\ref{eq02}), we have:
\begin{equation}\label{eq03}
  {P_{out} \over P_{in}} = {\kappa_R \over i(\omega_p-\omega)+\kappa_p + {g^2 \over i(\omega_m-\omega)+\kappa_m}}
\end{equation}
with $g=\sqrt{\omega_p\omega_M V_M/4V_c}$. Eq. (\ref{eq03}) is the fundamental equation for cavity magnonics and describes the anticrossing spectra in the strong coupling regime ($g>\kappa_p,\kappa_m$). For the coupling strength, taking $M_sV_M=N\mu_B$, where $\mu_B=\gamma\hbar$ is the magnetic moment for a single Bohr magneton and $N$ is the total number of spins, the expression for the coupling strength is:
\begin{equation}\label{eq04}
g=\gamma \sqrt{N\mu_0\hbar\omega_p \over 4V_c}
\end{equation}
Eq. (\ref{eq04}) defines one of the most important parameters in cavity magnonics which determines the excitation transduction bandwidth or the Rabi-like oscillation between magnons and photons. In addition, Eq. (\ref{eq04}) reveals that $g$ goes up when 1) $N$ becomes larger and 2) $V_c$ becomes smaller. For 1), it is straightforward to increase the volume of the magnetic sample in many cases. However the requirement for device miniaturization and integration requires 2) to obtain higher sensitivity to magnon excitations.

\begin{figure}[htb]
 \centering
 \includegraphics[width=3.0 in]{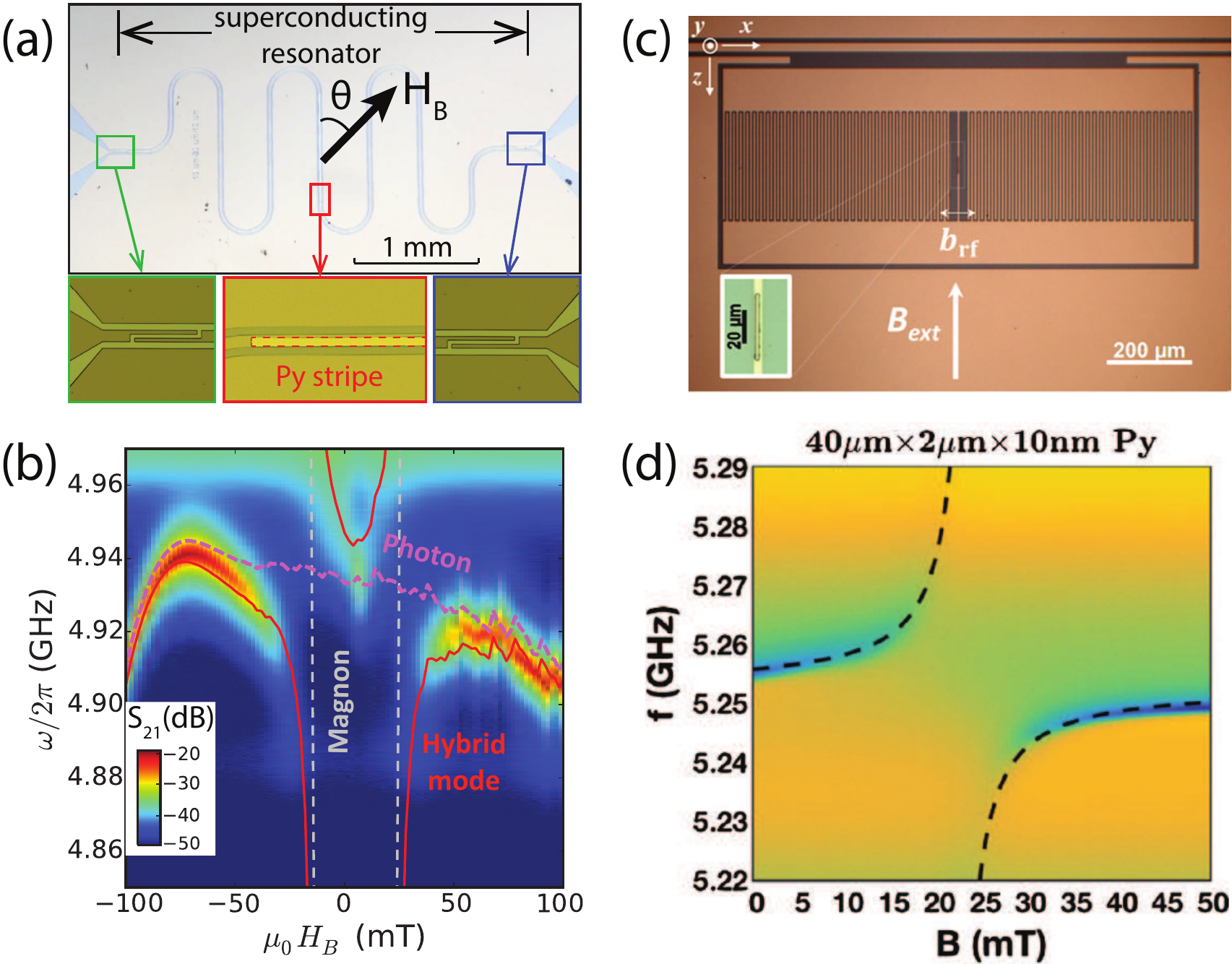}
 \caption{All-on-chip magnon-photon hybrid system with superconducting resonators. (a) Coplanar superconducting resonator coupled to on-chip Py device. (b) Mode anti-crossing spectra for a 900~$\mu$m $\times$ 14~$\mu$m $\times$ 30~nm Py device, measured from (a). (c) Lumped-element coplanar superconducting resonator coupled to on-chip Py device (d) Mode anti-crossing spectra for a 40~$\mu$m $\times$ 2~$\mu$m $\times$ 10~nm Py device, measured from (c). Adapted from Refs. \cite{LiPRL2019_magnon} and \cite{HouPRL2019}.}
 \label{fig_superconducting_resonator}
\end{figure}

\section{All-on-chip platforms for hybrid magnonics}

All-on-chip platforms for magnon-photon hybrid systems are desired for pushing towards microwave circuit applications and device integration. However, it is a dilemma that in many cases large volumes of magnetic samples are needed for strong coupling. The most commonly studied model system consists of a yttrium iron garnet (YIG) single crystal placed inside a 3D microwave cavity \cite{TabuchiPRL2014,ZhangPRL2014,GoryachevPRApplied2014,BaiPRL2015,
ZhangnpjQuantumInfo2015,LambertJAP2015,MaierFlaigPRB2016,KostylevAPL2016,BoventerPRB2018}, where YIG is a ferrimagnetic material that has the lowest known magnetic damping coefficient down to $10^{-5}$. Up to gigahertz coupling strength can be reached for large enough YIG crystals \cite{ZhangPRL2014,GoryachevPRApplied2014,KostylevAPL2016}. However, due to the macroscopic effective volume of the cavity, the coupling strength per Bohr magneton, defined as $g_0 = g/\sqrt{N}$, is usually small, much lower than 1 Hz. This also requires a macroscopic volume of the YIG crystal, typically with a diameter of 1 mm, in order to obtain strong coupling. Planar microwave resonators such as split-ring resonators (SRRs) \cite{StenningOE2013,BhoiJAP2014,KaurAPL2016,KlinglerAPL2016,BhoiSREP2017,ZhangDongshanJPD2017} and stripline resonators \cite{CastelIEEE2017,YaoNComm2017,YangPRApplied2019} have been also explored for their better on-chip integration. Because of the lower dimensions, planar resonators usually have smaller effective volumes and thus larger $g_0$ compared with 3D cavities \cite{BhoiSREP2017}. However, due to much thinner conducting lines compared with thick 3D cavity walls, the resistive loss is overwhelming and the quality factors of the planar resonators are quite low, usually less than 100, which makes it hard to extend the cooperativity.

Superconducting coplanar resonators possess both planar geometry and high quality factor. In addition, a high quality factor can be maintained with very small effective volume, thus providing high spin sensitivity for device miniaturization and on-chip integrations. Recently, all-on-chip magnon-photon hybrid systems have been demonstrated \cite{CMHuPhysics2019} in the strong-coupling regime by Li \textit{et al.} \cite{LiPRL2019_magnon} and Hou \textit{et al.} \cite{HouPRL2019}. In both works, the magnon system consists of a Ni$_{80}$Fe$_{20}$ (Py) thin-film stripe fabricated directly on top of the signal line of the superconducting resonator, which is a half-wavelength coplanar cavity with a signal line width of 20 $\mu$m [Fig. \ref{fig_superconducting_resonator} (a)]. The effective volume of the resonator is $V_m=0.0051$ mm$^3$, much smaller than for 3D cavities. This enables a very high sensitivity of $g_0/2\pi=26.7$ Hz. The choice of using a Py device is mainly for the convenience of fabrication and thin film deposition, and Py has a higher magnetization (1 T) compared with YIG (0.2 T). Since all the spins in the Py stripe are in good proximity to the signal lines, they all experience optimal coupling to the microwave photons. A coupling strength of 152 MHz is achieved with a 900 $\mu$m$\times 14$ $\mu$m$\times 30$ nm stripe for Li \textit{et al.} \cite{LiPRL2019_magnon} and for Hou \textit{et al.} \cite{HouPRL2019} the coupling stength is 171 MHz for 2000 $\mu$m$\times 8$ $\mu$m$\times 50$ nm. The coupling strength can be easily extended by increasing the size of the Py stripe, which has been demonstrated in both works.

For Li \textit{et al.}'s work \cite{LiPRL2019_magnon}, the superconductor is NbN and has a high superconducting transition temperature of 14 K. The NbN films were deposited at room temperature with an additional ion-beam gun to assist the reaction between Nb and N$_2$ during deposition \cite{PolakovicAPLMaterials2018}, making NbN promising for CMOS-compatible processing. At 1.5 K where the circuit is characterized, a quality factor of $Q=7600$ for the unloaded resonator and $Q=2500$ for the loaded resonator are measured, the latter corresponding to a photon damping rate of $\kappa_p/2\pi=2.0$ MHz. Similar values have been also achieved by Hou \textit{et al.} \cite{HouPRL2019}. The quality factor is much higher than non-superconducting planar cavities. Furthermore, the loss in superconducting resonators is dominated by two-level system fluctuations, which can be suppressed at even lower temperatures \cite{CarterAPL2019}.

The achievement of strong coupling is represented by the clear mode anticrossing between the magnon mode and the photon mode, as shown in Fig. \ref{fig_superconducting_resonator} (b). A cooperativity of $C=68$ from Ref. \cite{LiPRL2019_magnon} and 160 from Ref. \cite{HouPRL2019} have been reported and show that both cases are in the strong coupling regime. The main bottleneck for increasing the cooperativity is the damping of Py, reported as $\alpha=0.017$ for the Py stripe \cite{LiPRL2019_magnon}. Although YIG can serve as an alternative for improved damping \cite{HueblPRL2013,MorrisSREP2017,McKenziePRB2019}, it still remains a challenge to integrate YIG thin film devices with superconducting resonators, mainly due to the Gd$_3$Ga$_5$O$_{12}$ (GGG) substrates needed for the epitaxial growth of YIG films. At cryogenic temperature, the high microwave losses at cryogenic temperatures in GGG  \cite{MihalceanuPRB2018,LiuPRB2018,KosenAPLMaterials2019} will reduce the quality factor of the superconducting resonator and undermine the coherence of excitations.

Besides half-wavelength superconducting resonator as shown in Fig. \ref{fig_superconducting_resonator}(a), different resonator designs can yield even smaller effective volume for higher sensitivity of magnon excitation. For example, quarter-wavelength design can cut the total volume of half-wavelength resonator in half \cite{CarterAPL2019,MandalarXiv2020}. Moreover, using the design of lumped-element \textit{LC} resonators which are based on planar capacitor (\textit{C}) and inductor (\textit{L}) \cite{McKenziePRB2019,HouPRL2019}, it is possible to further reduce the effective volume from the $L$ component and further increase $g_0$. Shown in Fig. \ref{fig_superconducting_resonator}(c), Hou \textit{et al.} \cite{HouPRL2019} also demonstrated strong magnon-photon coupling with $g_0/2\pi=263$ Hz, which is an order of magnitude larger than the value with the CPW design.

For the future perspective, we anticipate on-chip magnon-photon hybrid systems to become one of the main-stream platforms for studying coherent magnonic interactions. In particular, many new device concepts that have been developed with YIG crystals, as will be discussed in the next section, can be implanted into an on-chip geometry. As the majority of ferrite-based microwave components are bulky and inconvenient for on-chip integration, utilizing strong coupling and frequency tunability of magnons could lead to much more compacted and efficient microwave controllers or sensors as well as for magnonic logic circuits. Moreover, coherent interaction of on-chip magnonic devices with superconducting microwave circuits can lead to applications in quantum information and sensing.

\section{New properties in coherent magnonics}

With the booming of various magnon hybrid systems, new physics are being discovered by engineering the parameters and the coherence of the hybrid components such as the microwave photons and magnons. In addition, many device concepts can be borrowed from the community of photonics and quantum information, where coherent information processing has been already well implemented. Below we discuss a few examples to show the impact of coherent magnon dynamics in both physics and engineering.

\subsection{Coherent spin pumping}

\begin{figure}[htb]
 \centering
 \includegraphics[width=3.2 in]{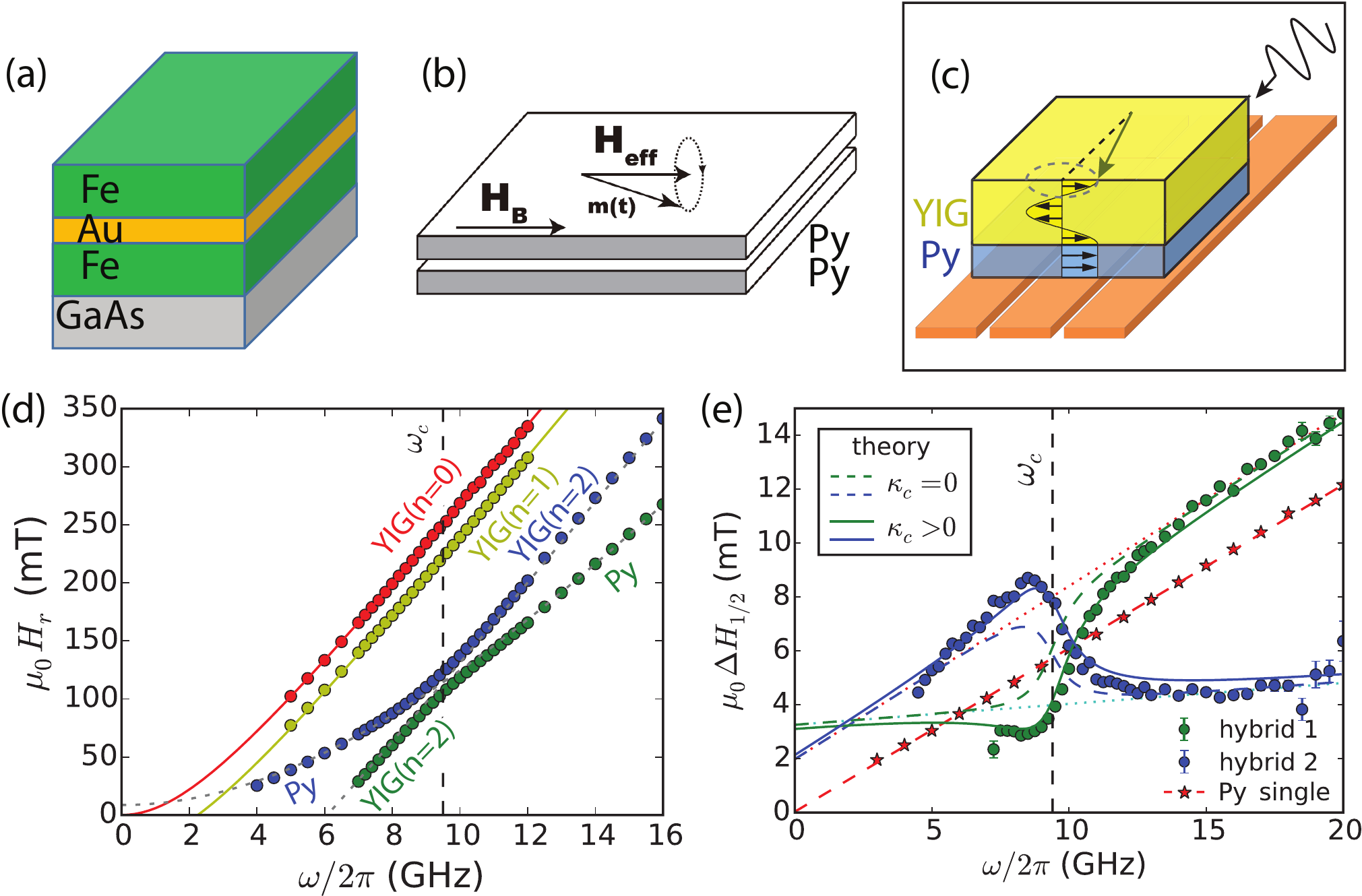}
 \caption{Coherent spin pumping in magnon-magnon hybrid systems. (a-c) Sample structures of two dynamically coupled magnetic layers with (a) asymmetric Fe/Au/Fe trilayer, (b) symmetric Py/Ru/Py trilayer and (c) YIG/Py bilayer, with mode asymmetry realized by (a) magnetocrystalline anisotropy, (b) RKKY interaction and (c) interfacial exchange coupling. (d) Mode anti-crossing for a YIG(100~nm)/Py(9~nm) bilayer as shown in (c). (e) Linewidth evolution for a YIG(100~nm)/Py(7.5~nm) bilayer, showing the effect of coherent spin pumping and dissipative coupling. Adapted from Refs. \cite{HeinrichPRL2003}, \cite{YangAPL2016} and \cite{LiPRL2020_YIGPy}.}
 \label{fig_csp}
\end{figure}

Spin pumping describes the process of pumping spin angular momentum from precessing magnetization \cite{tserkovnyakPRL2002,TserkovnyakRMP2005}. In spintronics, spin pumping is an important phenomenon because it is widely used for determining spin-to-charge conversion efficiency \cite{MosendzPRL2010,CzeschkaPRL2011,SanchezNcomm2013,BaiPRL2013,HahnPRB2013,ZhangWeiPRL2014,ShiomiPRL2014,NanPRB2015,JungfleischPRB2015,LesneNatureMater2016,RiddifordAPL2019,HanWeiNatureMater2020}, spin mixing conductance and spin diffusion lengths \cite{GhoshPRL2012,ShawPRB2012,SanchezPRL2014,WangPRL2014} in diverse material systems. The pumped spin angular momentum per area, or spin current, can be expressed as $I_{sp}/S=(\hbar/4\pi)\sigma^{\uparrow\downarrow}\mathbf{m}\times d\mathbf{m}/dt$, where $\mathbf{m}$ is the macrospin unit vector, $\sigma^{\uparrow\downarrow}$ is the spin mixing conductance with the unit of numbers of channels per unit area (note that in literature spin mixing conductance is conventionally represented by $g^{\uparrow\downarrow}$ but here we use $\sigma^{\uparrow\downarrow}$ instead to distinguish from the coupling strength $g$). This definition suggests that spin pumping is a \textit{coherent} process, because $I_{sp}$ always keeps a constant phase with $\mathbf{m}$ and has the same frequency as the dynamics of magnetization. Confirming the coherence of spin pumping is thus a fundamental question to address in magnetization dynamics and also important for coherent interactions of magnons with other excitations in magnon hybrid systems.

Magnon-magnon hybrid systems provide a unique platform to explore the coherence of spin pumping, because the phase of the magnon excitations can be arbitrarily adjusted by changing the frequency detuning of the two magnon system. Here the pumped spin currents from the two magnetic layers will interfere with each other: when the magnetization precessions of the two layers are in-phase, the pumped spin currents cancel out, leading to suppressed damping; when the precessions are out-of-phase, the pumped spin currents add up, leading to enhanced damping. The seminal work by Heinrich \textit{et al.} \cite{HeinrichPRL2003} has realized the mode crossing and in-phase interference of mutually pumped spin current in Fe/Au/Fe trilayers by utilizing the magnetocrystalline anisotropy of the epitaxial Fe. Then Yang \textit{et al.} \cite{YangAPL2016} showed clear damping difference between the acoustic (in-phase) and optic (out-of-phase) modes in a symmetric Py/Ru/Py structure. Owing to the Ruderman-Kittel-Kasuya-Yosida (RKKY) coupling via the Ru layer, the frequencies of the acoustic and optic modes are well separated, which facilitates the damping extraction. The issue is that the frequency detuning of the two Py layers is always zero and cannot be modified, thus only the zero (in-phase) and $\pi$ (out-of-phase) phase modes can be excited. It is worth noting that when the two magnetic layers are not degenerate in frequency, the pumped spin current from the resonance of one layer can drive the other layer into motion, and a phase offset of 90 degrees has been found between the two layers \cite{WoltersdorfPRL2007} which validates the phase relation between the magnetization dynamics and its spin pumping. However, interference between pumped spin current and magnetization dynamics would be a direct evidence of coherence and is more interesting in coherent information processing.

Recently, Li \textit{et al.} \cite{LiPRL2020_YIGPy} have shown the coherence of spin pumping in a strongly coupled magnon-magnon hybrid system with YIG/Py thin-film bilayers. The coherent coupling is provided by the interfacial exchange between YIG and Py. By exciting the perpendicular standing spin wave (PSSW) modes in YIG, mode intersection can be realized and engineered \cite{KlinglerPRL2018,ChenPRL2018,QinSREP2018}. In addition, the frequency detuning can be controlled by the external biasing field, leading to the observation of mode anticrossing. In the strong coupling regime, the two hybrid modes are well separated and their linewidths can be individually extracted. From the linewidth plot, it is clear that when the frequency detuning is zero at $\omega_c/2\pi=9.4$ GHz, one hybrid mode has a significantly larger linewidth than the other, similar to the observation of linewidth difference in Py/Ru/Py structure \cite{YangAPL2016}. This is due to the effect of spin pumping from the YIG/Py interface. The full evolution of the spin pumping linewidth enhancement, with the precession phase difference decided by the frequency detuning, can be nicely fitted to the theory, showing the coherence of the spin pumping. Moreover, the coherent nature of the dissipative spin pumping process leads to a dissipative coupling, or an imaginary component of $g$ in Eq. (\ref{eq03}), as compared to the real component of $g$ from interfacial exchange. The dissipative coupling is important for engineering the hybrid dynamics, with examples of level attraction which has in fact been observed already in the work from Heinrich \textit{et al.} \cite{HeinrichPRL2003}.

\subsection{Exceptional points}

In a two-level system, exceptional points are mathematically defined as the singularities where the two eigenvalues, and their corresponding eigenvectors, simultaneously coalesce \cite{HeissJPA2012,MiriScience2019}. In a magnon-photon hybrid system, the two eigenvalues of Eq. (\ref{eq01}) are:
\begin{align}\label{eq10}
  \omega_\pm = {\omega_m+\omega_p \over 2} - {\kappa_m+\kappa_p \over 2} \pm {1\over 2}\sqrt{(\Delta\omega-i\Delta\kappa)^2+4g^2}
\end{align}
where $\Delta\omega=\omega_m-\omega_p$ and $\Delta\kappa=\kappa_m-\kappa_p$. The exceptional point happens when $\Delta\omega=0$ and $\Delta\kappa = 2g$. At this condition, the square root term becomes zero and two eigenvalues become one. Expressed differently, at the exceptional points the Hamiltonian cannot be fully diagonized and only one eigen-solution remains. This peculiar property has prompt many studies of non-Hermitian systems and applications in enhanced sensing \cite{HodaeiNature2017,ChenNature2017}.
\cite{}
\begin{figure}[htb]
 \centering
 \includegraphics[width=3.0 in]{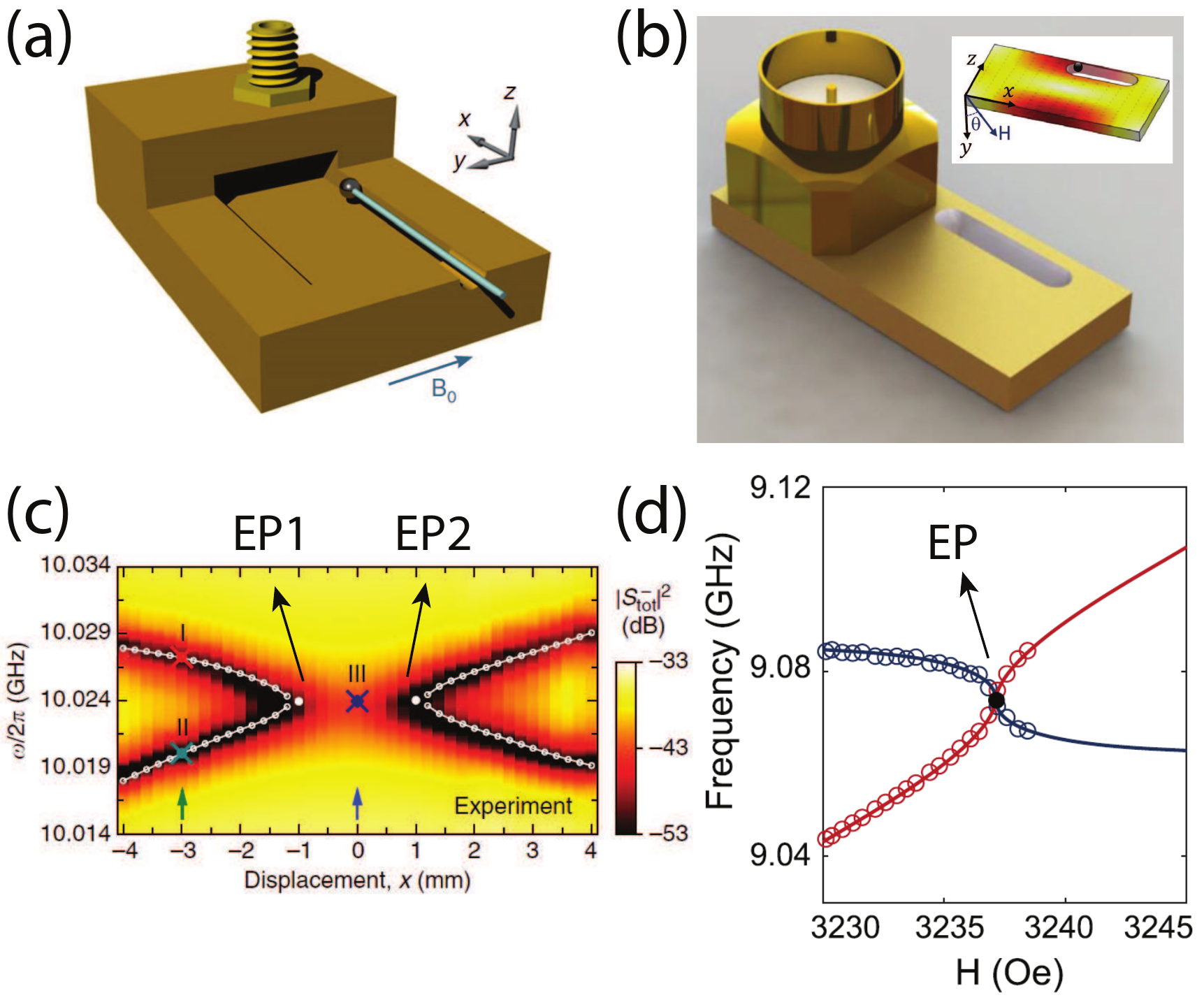}
 \caption{Realizing exceptional points with tunable magnon-photon coupling. (a-b) Two different microwave cavity structures for tunable magnon-photon coupling strength. (c) Demonstration of exceptional points by tuning the insertion depth of the YIG sphere for the geometry in (a). (d) Enhanced $d\omega/dH$ slope and sensitivity at one exceptional point for the geometry in (b). Adapted from Refs. \cite{ZhangDengkeNComm2017} and \cite{ZhangPRL2019}.}
 \label{fig_EP}
\end{figure}

The advantage of magnon-photon hybrid systems is that $\Delta\omega$, $\Delta\kappa$ and $g$ can be all tuned \textit{in-situ} in order to realize the exceptional point, which has been shown in two recent works \cite{ZhangDengkeNComm2017,ZhangPRL2019}. In each case, the magnon system is based on a single-crystal YIG sphere and the microwave cavity is defined by a 3D rectangles [\ref{fig_EP}(a-b)]. The YIG sphere is mounted on a movable rod and the position of the YIG sphere in the cavity can be modified in order to change $g$. Fig. \ref{fig_EP}(c) shows the measured microwave reflection of the cavity as a function of the YIG insertion depth $x$ at $\Delta\omega=0$, with $g$ tuned by $x$ as $g/2\pi=1.3|x|$ MHz/mm. When $|x|$ is large, the system is in the strong coupling regime and two dips in the spectrum represent the two split modes at the anticrossing gap minimum. As $|x|$ is decreased to 1.2 mm, the two dips merge into a single dip, which indicates the exceptional point. Further decreasing $|x|$ (and $g$) will lead to the intrinsic resonant absorption of the cavity with the magnon mode decoupled. The exceptional point also marks the transition from the weak-coupling regime to the magnetically induced transparancy (MIT) regime \cite{ZhangPRL2014}. Furthermore, in Ref. \cite{ZhangPRL2019} the value of $g$ can be tuned by $x$, $y$ and the field angle $\theta$, thus allowing the determination of an exceptional surface in the variable space and exploration for high-dimension non-Hermitian physics \cite{ZhangGuoQiangPRB2018,CaoPRB2019}. We also note that recently exceptional points between two coupled YIG spheres have been demonstrated mediated by a microwave cavity \cite{ZhaoJiePRApplied2020}. Here the two magnonic resonators are dissipatively coupled ($g$ is imaginary) so the exceptional points take place at a finite frequency detuning.

The main technical interest of exceptional points is that the sensitivity can be significantly enhanced \cite{ZhangGuoQiangPRB2018,CaoPRB2019,YuanPRL2020}. This can be demonstrated in Fig. \ref{fig_EP}(d), where the field dependence of the two hybrid modes shows a square-root behavior and with enhanced slope near the exceptional point \cite{ZhangPRL2019}. This is because in Eq. (\ref{eq10}) when $\Delta\kappa = 2g$ and with small frequency perturbation $\Delta\omega \ll \Delta \kappa$, the square-root term becomes $\pm \sqrt{i\Delta\kappa\Delta\omega}/2$, which exhibit a $\Delta\omega^{1/2}$ dependence and can generate a much larger change of $\omega_\pm$ for a given $\Delta\omega$ compared with the uncoupled regime ($g=0$, $\delta\omega_\pm=\pm\Delta\omega/2$). We also note the exploration of exceptional points in RKKY coupled magnon-magnon hybrid systems, where the coupling strength $g$ can be tuned by the thickness of the barrier of the synthetic antiferrimagnet \cite{LiuSciAdv2019,YuTianlinPRB2020}. This provides the all-thin-film based platform for exploring non-Hermitian physics and advanced sensing.

\subsection{Nonreciprocity}

\begin{figure*}[htb]
 \centering
 \includegraphics[width=5.5 in]{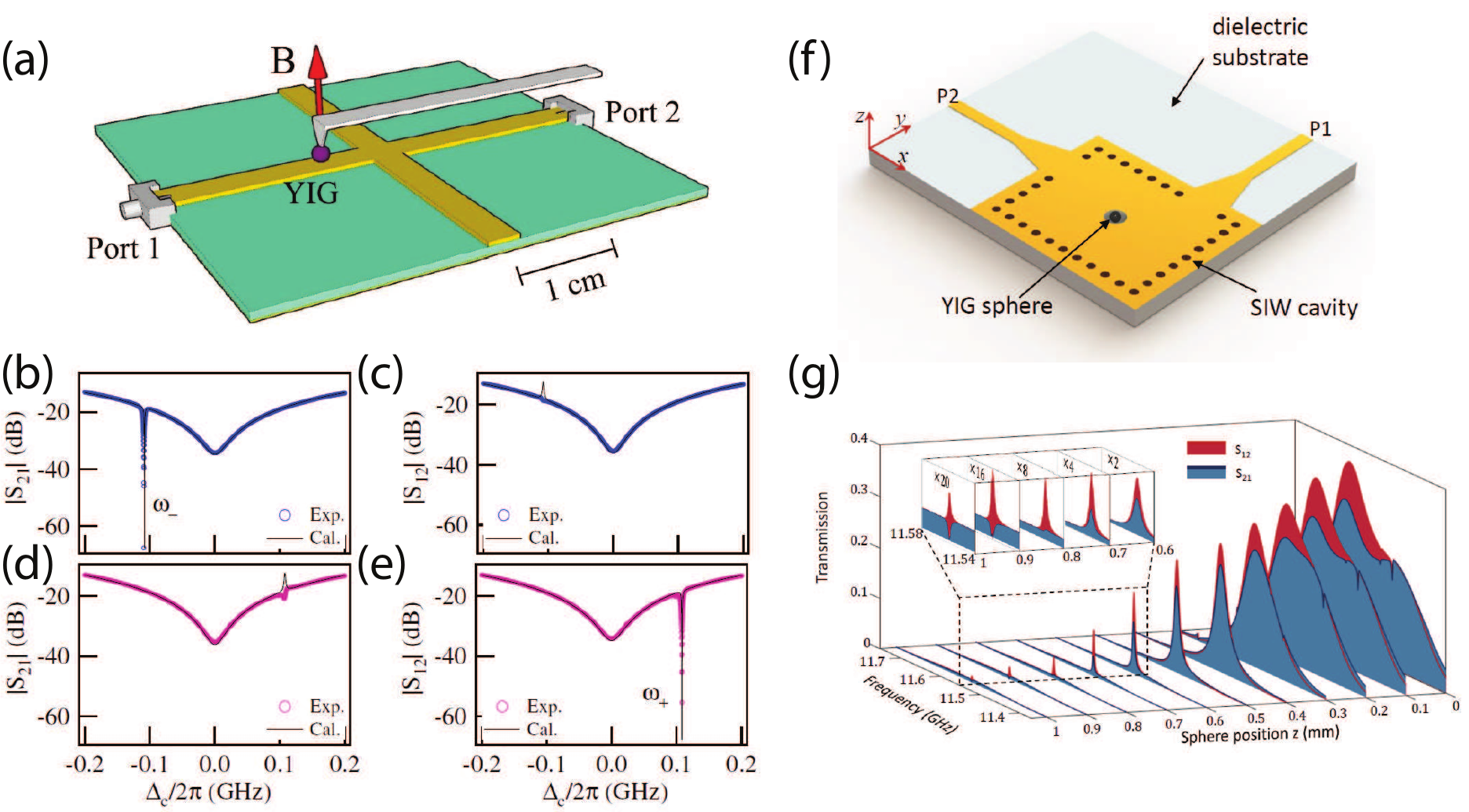}
 \caption{Realizing microwave nonreciprocity with engineered magnon-photon hybrid systems. (a) Microwave resonator design incorporating both traveling and standing waves, with the position of the YIG sphere adjusted by a 3D positioner.  (b-c) Nonreciprocal $|S_{21}|$ and $|S_{12}|$ measured from the circuit when the magnetic field is tuned to $\omega_m=\omega_c-2J\Gamma/\alpha$. Complete isolation valley is observe at $\omega_-$ for $|S_{21}|$, while no valley is observed for $|S_{12}|$. (d-e) Same for (b-c) except that the magnetic field is tuned to $\omega_m=\omega_c+2J\Gamma/\alpha$. No isolation valley is observe at $\omega_-$ for $|S_{21}|$, while complete isolation is observed for $|S_{12}|$. (f) Microwave resonator design with two circularly polarized degenerate modes. (g) Nonreciprocal $S_{21}$ and $S_{12}$ measured for different magnon-photon coupling strengths (YIG locations). At $z=0.9$ mm complete isolate valley is observed for $S_{21}$, corresponding to the destructive interference between the cavity mode and the crosstalk transmission mode. Adapted from Refs. \cite{WangPRL2019} and \cite{ZhangPRApplied2020}.}
 \label{fig_Nonreciprocity}
\end{figure*}

Nonreciprocity is an important concept that has found many applications in microwave and optical communications\cite{VerhagenNPhys2017,SounasNPhotonics2017,CalozPRApplied2018}, with examples of circulators and isolators. It is important to implement noise-free, phase preserving information processing and amplification especially in the quantum regime. To realize nonreciprocal signal transmission, one needs to break the time-reversal symmetry of the system. The traditional examples are microwave ferrite circulators and optical Faraday isolators, where the symmetry is broken by the magnetization vector of ferrites in the devices. Symmetry-breaking can be also artificially introduced by engineering the propagation paths of the electromagnetic waves for asymmetric phase accumulation, with recent progress in photonics \cite{SounasNPhotonics2017}, optomechanics \cite{ShenNPhotonics2016,RuesinkNComm2016,FangNPhys2017} and microwave circuits \cite{LecocqPRL2016,BernierNComm2017,PetersonPRX2017}.

Incorporating magnon-photon hybrid systems opens up new features in controlling nonreciprocity for coherent information processing with magnonics. Recently, two groups have demonstrated nonreciprocal microwave transmission by coherently coupling a circuit-board microwave cavity to a YIG sphere \cite{WangPRL2019,ZhangPRApplied2020}. In particular, nearly 100\% isolation can be achieved from the engineered cooperative dynamics of magnon and photon resonance. Compared with other schematics to create nonreciprocity, magnon hybrid systems are highly tunable in frequency, phase and coupling strengths, allowing for convenient engineering and implementation of desired nonreciprocal characteristics.

In the first work by Wang, \textit{et al.} \cite{WangPRL2019}, the microwave cavity is a half-wavelength microstrip resonator forming a cross line with the feeding microtrip transmission line, as shown in Fig. \ref{fig_Nonreciprocity}(a). A YIG sphere is placed adjacent to both the cavity and the microstrip line so that the magnon mode is coupled to both the standing-wave photons (cavity) and the propagating photons (microstrip). It has been shown that dissipative coupling between magnons and photons can be achieved when the two harmonic resonator systems are both coupled to the same traveling photons \cite{YuPRL2019,YangPRApplied2019,RaoPRB2020}. This is also known as reservoir engineering \cite{PoyatosPRL1996,MetelmannPRX2015,AluPRApplied2017}. With proper adjustment of the YIG sphere location, the coherent (real) and dissipative (imaginary) couplings can be tuned to be equal and in balance. In this situation, when the magnon and photon resonators have a certain frequency detuning, the damping of the magnon-dominating hybrid mode will be completely compensated with the energy pumping from the dissipative coupling. From the viewpoint of microwave circuits, a zero-dissipation resonator will absorb all the energy and completely block microwave transmission. This is reflected as a second inverse resonance peak with infinite depth [$\omega_-$ in Fig. \ref{fig_Nonreciprocity}(b), where $\omega_+$ corresponds to the cavity-dominating peak near $\Delta_c/2\pi=0$]. However, when the microwave propagating direction is reversed, the sign of the dissipative coupling will also be reversed. This will in turn reverse the sign of the frequency detuning between magnons and photons in order to reach the zero-damping condition, with the infinite-depth resonance peak ending up on the other side of the broad cavity peak [$\omega_+$ in Fig. \ref{fig_Nonreciprocity}(e)]. Thus nonreciprocal microwave transmission with perfect isolation is realized.

In the second work by Zhang, \textit{et al.} \cite{ZhangPRApplied2020}, the microwave cavity is a substrate integrated waveguide (SIW) cavity [Fig. \ref{fig_Nonreciprocity}(f)], which is a rectangular waveguide built into a circuit board with the sidewall defined by plated through hole vias. Compared with transmission line based microwave resonators, SIW cavities are more convenient for engineering the microwave profiles. In this case, the cavity can host circularly polarized eigenmodes, which are the combination of two linearly polarized eigenmodes with degenerate energy level by design. Furthermore, the two microwave ports are designed to excite only one circularly polarized modes. When the YIG sphere is saturated along the $z$ direction, the magnon excitations only couple to one helicity of the microwave mode, thus breaking the symmetry and creating the nonreciprocity. In the practical circuit, there is a finite crosstalk between the two port (meaning the transmission is nonzero even when the SIW cavity is not on resonance). The magnon-photon coupling mediated microwave transmission thus interfere with the crosstalk transmission constructively or destructively, leading to the complete isolation as the position of the YIG sphere is set to $z=0.9$ mm [Fig. \ref{fig_Nonreciprocity}(g)]. A similar effect based on a ring resonator has also been proposed \cite{ZhuNaPRA2020}.

In magnonics, nonreciprocity has also been extensively explored. The main mechanism is to create a difference in effective field for different magnon propagating directions, with examples of surface anisotropy \cite{GladiiPRB2016}, Dzyaloshinskii-Moriya interaction \cite{IguchiPRB2015,WangPRL2020} and dipolar field \cite{ChenPRB2019,GallardoPRApplied2019}. On the other hand, nonreciprocity based on magnon hybrid systems takes advantage of the phase and coupling engineering. In the above two examples, the magnon-photon coupling is employed to tune the amplitude and phase of the resonator photons, which interfere with the traveling microwave photons and create full isolation. This approach sacrifices bandwidth and gains isolation depth. For coherent information processing, circuits are usually running in narrow band, thus the tunability of magnons makes life much easier to match the bandwidth. Compared with conventional ferrite-based circulators, strong coupling between magnons and photons will significantly reduce the volume of magnetic materials needed for reaching isolation. We also note the development of surface acoustic wave (SAW) isolators based on magnon-phonon coupling \cite{LewisAPL1972,SasakiPRB2017,VerbaPRApplied2018}, where the nonreciprocal effects in magnonics are utilized to create nonreciprocity in SAWs when coupled to magnons. Similar to magnon-photon coupling, strong magnon-phonon coupling have also been proposed to produce deep isolation \cite{VerbaPRApplied2018}. This will benefit from the low damping and high-efficiency energy excitation of SAWs for microwave applications. However, one of the challenges remains to engineer strong coupling between magnons and SAW, which depends sensitively on the sample processing and microstructure \cite{ZhaoPRApplied2020}.

\section{New hybrid magnonic platforms for remote information transduction}

\begin{figure}[htb]
 \centering
 \includegraphics[width=3.0 in]{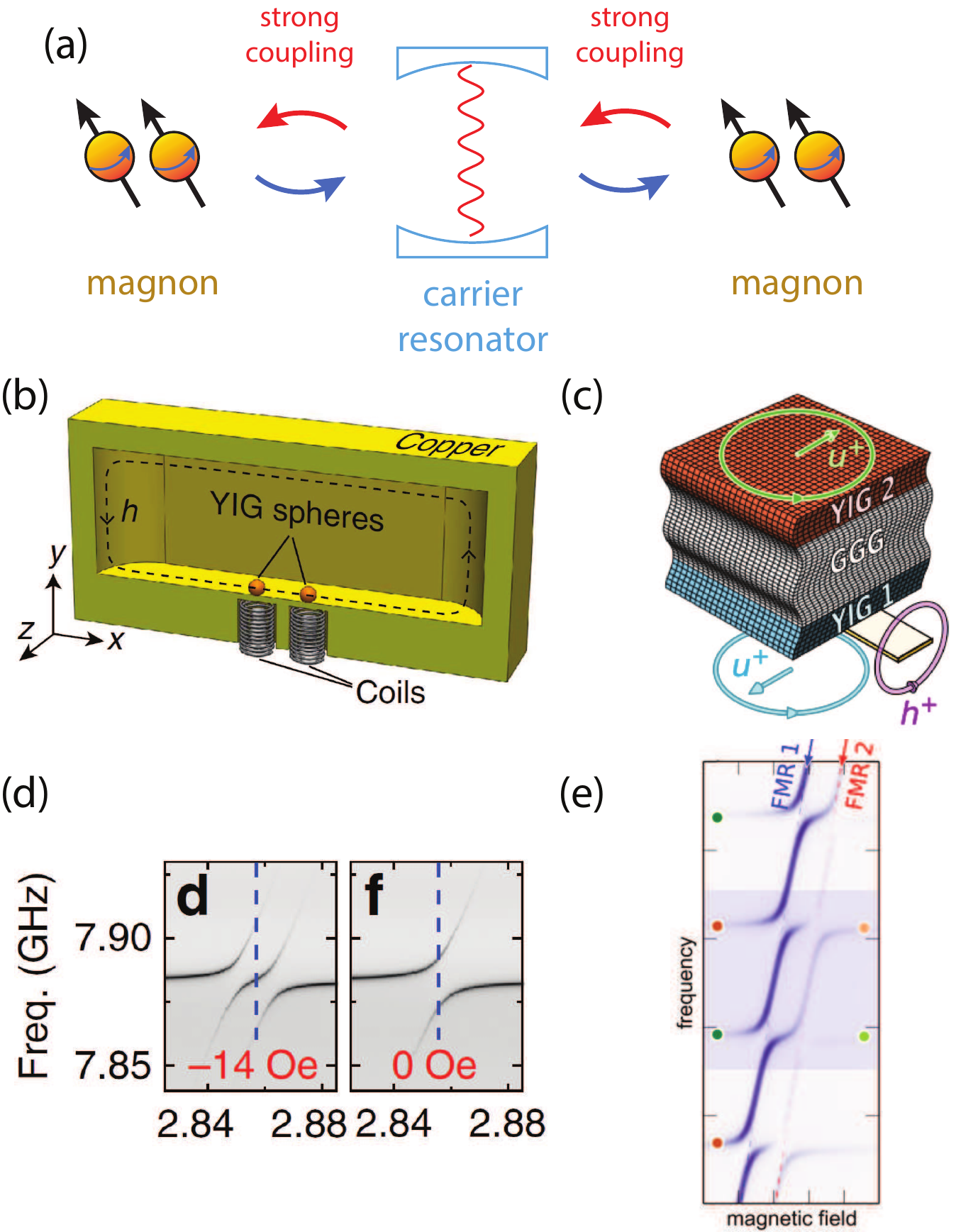}
 \caption{(a) Remote transduction of coherent magnon excitations in magnon hybrid systems. (b) Cavity photon mediated remote coupling between two YIG spheres. Two coils are used to introduce frequency detuning between the two YIG spheres. (c) Cavity reflection spectra for a field detuning of -14 Oe and 0 Oe. (d) Phonon mediated remote coupling between two YIG films by the GGG substrate. (e) Ferromagnetic resonance spectra of the YIG/GGG/YIG trilayer. Adapted from Refs. \cite{ZhangNComm2015} and \cite{AnPRB2020}.}
 \label{fig_transducer}
\end{figure}

Magnons can couple to many excitations that have played important roles in quantum information, such as microwave photons, optical photons, phonons, superconducting qubits and NV centers, making magnons highly relevant for quantum information transduction \cite{LachanceQuirionAPEx2019,KusminskiyarXiv2019}. As transducers, magnons have been employed to mediate the coupling between microwave photons and phonons \cite{ZhangScienceAdv2016,LiJiePRL2018}, as well as microwave and optical photons \cite{HaighPRA2015,OsadaPRL2016,ZhangPRL2016,HaighPRL2016}. For the application in magnonics, dynamic coupling between distant magnetic elements has become an emerging topic for magnon-based coherent information processing \cite{ZhangNComm2015,BaiPRL2017,ChenPRL2018,AnPRB2020,RuckriegelPRL2020}. While the role of transducer has been extensively reviewed elsewhere \cite{LachanceQuirionAPEx2019,KusminskiyarXiv2019}, here we briefly discuss the aspect of magnonics and take a few schematics as examples, especially for the purpose of coherent transduction between remote systems which is essential for scalable integration [Fig. \ref{fig_transducer}(a)].

\textbf{Cavity mediated remote transduction.} Strong coupling between microwave cavity and magnons can be used to coherently couple multiple magnetic systems with the information exchanged via microwave photons. One example is the coherent coupling between two YIG spheres in a 3D cavity \cite{ZhangNComm2015}, as shown in Fig. \ref{fig_transducer}(b). The key ingredient is that a pair of coils are placed beside the two YIG sphere, which can be used to control their field (frequency) detuning. Fig. \ref{fig_transducer}(d) shows two characteristic mode anticrossing for a field detuning of -14 Oe and 0 Oe. For a detuning of -14 Oe, two clear avoided crosses with the gap $2(g/2\pi)=13.4$ MHz are observed corresponding to the strong coupling between the microwave cavity and each YIG sphere. When the detuning is set to zero, only one avoided cross is observed and the anticrossing gap is larger than the case of individual YIG spheres. What happens is that the degenerate magnon modes of the two YIG spheres have formed new collective modes as in-phase and out-of-phase dynamics. The locations of the YIG spheres are selected such that the microwave fields of the cavity photons are identical and in-phase. Thus the microwave photons only couple to the in-phase mode and are decoupled from the out-of-phase mode. The in-phase mode shows an gap of $2\sqrt{2}(g/2\pi)=18.9$ MHz, $\sqrt{2}$ times larger than the gap for a single YIG sphere. This is because the total volume of two YIG spheres is twice the value of a single sphere, thus the number of spins also doubles.

Let's imagine how the two YIG spheres communicate in the triple-zero-detuning case ($\omega_\text{YIG1}=\omega_\text{YIG2}=\omega_\text{cavity}$). If initially the excitation resides in YIG 1 with an amplitude of $A$ for the transverse magnetization motion, the total excitations of the two YIG spheres can be viewed as the superposition of $<A/2$,$A/2>$ and $<A/2$,-$A/2>$, which correspond to the in-phase and out-of-phase modes, respectively. The in-phase mode will start its Rabi-like oscillation with cavity photon mode with a bandwidth of $\sqrt{2}g$. The out-of-phase mode, which does not interact with either the in-phase or the cavity mode, will remain as the eigenmode. Thus we will observe that the excitation in YIG 1 oscillates between $A$ and $A/2$, and the excitation in YIG 2 oscillations between 0 and $-A/2$, with a period of $2\pi/\sqrt{2}g$. Similar cavity-mediated remote magnon-magnon coupling have also been shown with different efforts \cite{LambertPRA2016,BaiPRL2017,RameshtiPRB2018,XuPengChaoPRB2019,GrigoryanPRB2019,ZhaoJiePRApplied2020}, including utilizing spin pumping \cite{BaiPRL2017} and in the dispersive regime \cite{LambertPRA2016,XuPengChaoPRB2019}.

In the same work \cite{ZhangNComm2015}, it has been pointed out that the out-of-phase mode acts as the dark mode and may have long coherence time because it is isolated from the hybrid system. To be able to read the excitation information from the dark mode, the system can be engineered such that a small frequency detuning $\Delta\omega$ is set between the two YIG sphere. In the absence of the microwave cavity, the in-phase and out-of-phase modes are no longer degenerate and will hop from one to the other. In the presence of the microwave cavity, only the in-phase mode couples to the cavity photons. Thus the cavity output signal will exhibit a periodic oscillation with time, and the period is determined by $\Delta\omega$, as the information is repeatedly stored into and retrieved from the dark mode. Zhang \textit{et al.} has demonstrated this process with up to eight YIG spheres \cite{ZhangNComm2015}.

\textbf{Phonon-mediated remote transduction.} Coherent interactions betweens magnons and phonons result in their hybrid excitations known as magnon polarons \cite{KamraPRB2015,ShenPRL2015,KikkawaPRL2016,FlebusPRB2017,BozhkoPRL2017,HayashiPRL2018,HolandaNPhys2018}. They are different from incoherent magnon-phonon interactions as studied in spin Seebeck effect. In particular, phonons have been extensively explored in coherent information processing such as in optomechanics and quantum information; coupling with magnons can incorporate the great tunability of magnons into the phonon systems. Recently, An, \textit{et al.} \cite{AnPRB2020} have explore phonon-mediated remote magnon interaction in a YIG(200 nm)/GGG(500 $\mu$m)/YIG(200 nm) structure [Fig. \ref{fig_transducer}(c)]. The two YIG layers are grown by liquid phase epitaxy on a two-side-polished GGG substrate. For a one-side YIG sample, multiple mode anticrossings are observed between the YIG Kittel mode and the perpendicular standing acoustic waves ($n\sim 1400$) of the GGG substrate. Note that the acoustic wave has 1-cm half-decay length and a coherence time of 2 $\mu$s at 5 GHz, much better than what magnons can achieve. For the two sided YIG sample, the mode anticrossing spectra is shown in Fig. \ref{fig_transducer}(e). A small frequency detuning between the two YIG layers comes from the variation of magnetization. Similar to the previous example, there are in-phase and out-of-phase hybrid modes for the two YIG layers. The difference is that the phonon-mediated coupling favors the two hybrid modes alternatively for the excitation of the odd or even phonon modes. When the FMR of one YIG layer is excited, phonon-mediated coupling will drive the other YIG layer to precess in-phase or out-of-phase with respect to the excited YIG layer. Thus the total microwave absorption will be increased or decreased, respectively, as a result of constructive or destructive interference between the two YIG layers. This interference is measured as alternating FMR amplitudes along the Kittel dispersion line [labeled as FMR 1 and FMR 2 in Fig. \ref{fig_transducer}(e)].

\section{Future outlooks}

The development of magnon hybrid systems, including their advantages of on-chip integration, exploiting new properties in coherent magnonics and remote information transduction, will bring about new interdisciplinary opportunities in coherent information processing. Below we discuss the role of magnons in two emerging opportunities: quantum information and magnonics.

\subsection{Quantum information}

\begin{figure}[htb]
 \centering
 \includegraphics[width=3.0 in]{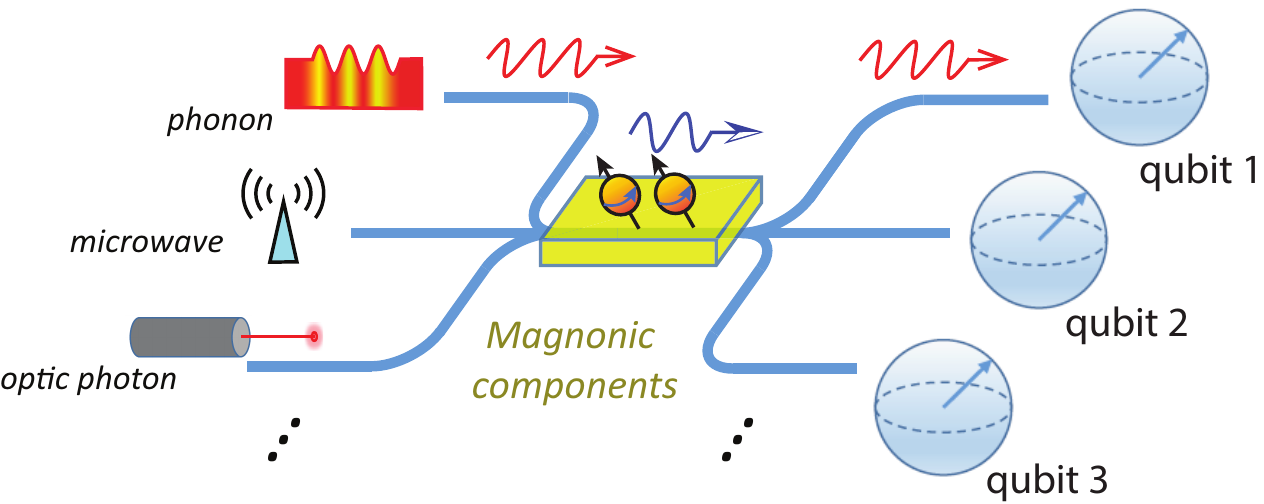}
 \caption{Diagram of magnonic components acting as on-chip quantum transducers and controllers for different qubits.}
 \label{fig_QIS}
\end{figure}

With the demonstration of single-magnon interaction with a qubit \cite{TabuchiScience2015,LachanceScienceAdvan2017,LachanceQuirionScience2020}, quantum magnonics has become a realistic topic. The main potential of magnons in quantum information is to act as an intermediate information transfer station for quantum transduction, coherent information control and entanglement generation (Fig. \ref{fig_QIS}). In particular, the convenience of tunability and on-chip integration are keys for applications in quantum information.

We anticipate two future platforms for addressing qubits with on-chip geometry and scalability. The first platform is the hybrid system between magnonic devices and superconducting microwave circuits. With planar superconducting microwave resonators, the novel functionality from cavity spintronics such as nonreciprocal transmission, isolation and enhanced sensitivity should be achievable in an all-on-chip architecture. The damping rate of single-crystal YIG can be as low as 1 MHz at a frequency of 5 GHz \cite{TabuchiPRL2014,KosenAPLMaterials2019,PfirrmannPRResearch2019} along with a sub-GHz magnon-photon coupling strength, which ensures hundreds of coherent Rabi-like oscillation cycles. However, integrating YIG devices on planar superconducting circuits would be a challenge. This is because that the ideal magnetic material, YIG thin films, needs to be grown on GGG substrate in order to maintain its low damping property, but at cryogenic temperature GGG has complex microwave property and exhibits large losses, which will undermine the coherence of all microwave excitations. To realize GGG-free YIG thin film devices, one way is to fabricate free-standing YIG thin-film devices \cite{SeoAPL2017,HeyrothPRApplied2019} or cut YIG crystals into small pieces and transfer them onto superconducting microwave circuits. GGG-free YIG growth such as on silicon \cite{FanPRApplied2020} may provide an alternative solution. New low-damping magnetic materials such as CoFe \cite{SchoenNPhys2016,LeeNComm2018,LiPRL2019_CoFe}, complex magnetic insulators \cite{EmoriNanoLett2018} and organic ferrimagnets \cite{FransonAPLMaterials2019} could provide new ideas for device implementation at cryogenic temperature. Besides microwave photons, magnon-mediated coherent transductions of optical photons and phonons provide new quantum interfaces and research topics such as cavity optomagnonics \cite{KusminskiyarXiv2019} and cavity magnetomechanics \cite{ZhangScienceAdv2016}. We also note the efforts of optical waveguides fabricated from magnonic materials, compared with bulk magnetic crystals, for on-chip microwave-to-optics conversion \cite{RossNPhotonics2011,ZhuarXiv2020}. Furthermore, we point out the electrical excitations of SAWs have been recently incorporated in quantum information \cite{SatzingerNature2018,WhiteleyNphys2019,BienfaitScience2019,DelsingRoadmap2019}. Studying magnon-photon interactions with SAWs \cite{WeilerPRL2011,BhuktareSREP2017,KobayashiPRL2017,ZhaoPRApplied2020} will benefit from convenient bandwidth engineering \cite{FuNComm2019} compared with exciting the bulk mechanical resonance modes.

The second emerging platform is the coherent interaction with spin qubits such as NV centers \cite{AndrichnpjQuantumInf2017}. The underlying physics is that propagating spin waves exert an effective rf field to a proximal NV center at the surface of the magnonic waveguide (5-10 nm away), which excites the NV center and triggers the optically detected magnetic resonance (ODMR). Because of the excitation confinement in magnetic media, magnon excitations can propagate for long distance, by a range of $\sim 100$ $\mu$m in low-damping YIG, and are shown to provide a few hundreds times stronger rf field than then radiation from the microwave antenna \cite{AndrichnpjQuantumInf2017}. This capability can be applied to selectively addressing remote spin qubits by utilizing the controllability of magnons, acting as a programmable microwave control bus. The interaction between magnons and NV centers has been also applied to NV center magnetometry owing to the ultrahigh sensitivity and spatial resolution \cite{GrossNature2017,DuScience2017,LeewongNanoLett2020,BertelliarXiv2020,ZhouarXiv2020}. Furthermore, all the experiments can be realized at room temperature, making YIG an ideal low-damping material for carrying magnon excitations. Another proposal is to implement long-distance entanglement of spin qubits mediated by propagating magnons \cite{TrifunovicPRX2013}. The operation time of the two-qubit gate, when the spin qubits are 25 nm away from the thin-film ferromagnetic transducer, is estimated to be a few tens of nanoseconds, which is much faster than the decoherence time of tens of $\mu$s for NV centers. One challenge will be the on-chip integration of high-quality YIG and diamond systems, which requires material engineering for compatible magnon-NV hybrid platforms.
\cite{}
\begin{figure}[htb]
 \centering
 \includegraphics[width=3.0 in]{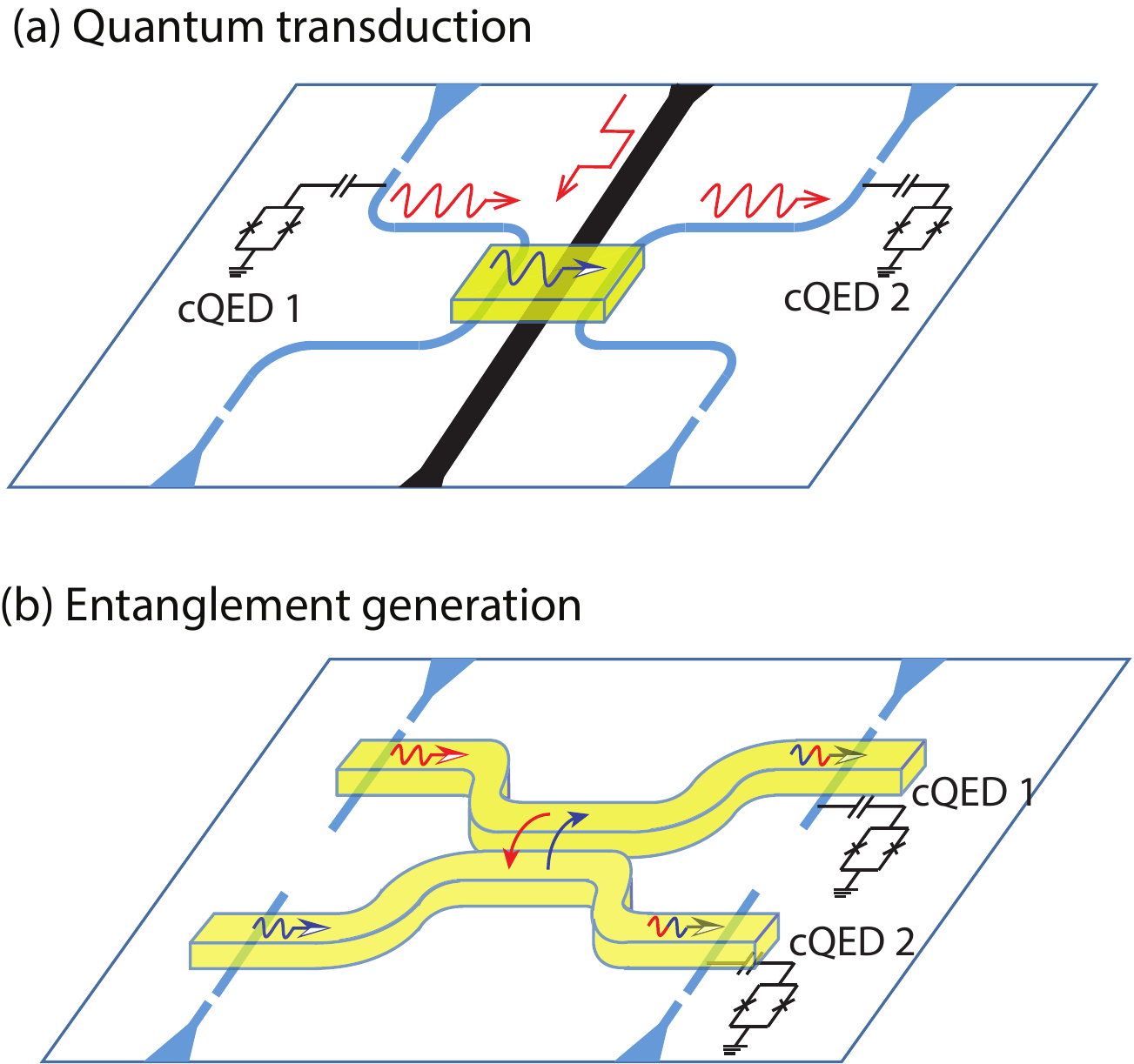}
 \caption{Proposed functionality in quantum information with magnon hybrid system. (a) Quantum transduction with magnons between two cQED systems. (b) Magnonic hybrid splitter for entanglement generation. Blue lines denote coplanar microwave waveguide. Yellow blocks denote magnetic devices and circuits. Black circuits denote superconducting qubits. The broad black stripe in (a) denotes controlling microwave electrodes for tuning the magnon frequency. }
 \label{fig_QIS_design}
\end{figure}


Figs. \ref{fig_QIS_design}(a) and (b) show the illustrations for two proposed functionality, quantum transduction and entanglement generation, respectively, with on-chip magnonics and superconducting circuits. In Fig. \ref{fig_QIS_design}(a), a magnetic element overlaps and couples with two circuit quantum electrodynamics (cQED) systems (each cQED system contains a superconducting microwave cavity and a qubit which are strongly coupled). This leads to strong magnon-photon coupling on both side, whilst cQED systems do not interact with each other directly. Quantum excitation from one side of cQED can be transferred to the other side mediated by magnons. Due to the frequency tunability of magnons, the quantum transduction can be turned on or off by setting the magnon frequency as degenerate with or far away from the frequency of the two cQED systems. This magnon frequency tuning process can be realized with high bandwidth by using an adjacent current-carrying line as shown in Fig. \ref{fig_QIS_design}(a).

Quasi-classically, the process of magnon-mediated coherent information transfer in such non-stationary system can be described by a system of coupled equations for the complex amplitudes $c_{1,2}(t)$ and $c_m(t)$ of the excitations in the superconducting microwave cavities and magnonic resonator, respectively \cite{TrevillianIntermag2020}:

\begin{subequations}\label{eq:nonstationary}
 \begin{eqnarray}
 \frac{dc_1}{dt} &=& - i\omega_1 c_1 - \kappa_1 c_1 -i g_{1m} c_m \,,\\
 \frac{dc_2}{dt} &=& - i\omega_2 c_2 - \kappa_2 c_2 -i g_{2m} c_m \,,\\
 \frac{dc_m}{dt} &=& - i\omega_m(t) c_m - \kappa_m c_m -i g_{1m} c_1 - i g_{2m} c_2
 \end{eqnarray}
\end{subequations}

Here $\omega_{1,2}$ are the eigen-frequencies of the cavities, $\omega_m(t)$ is the time-dependent magnon frequency, $\kappa_{1,2}$ are the decoherence rates for the two cavities, and $g_{im}$ are the coupling rates between cavity $i$ and magnonic resonator.

\begin{figure}[htb]
 \centering
 \includegraphics[width=3.0 in]{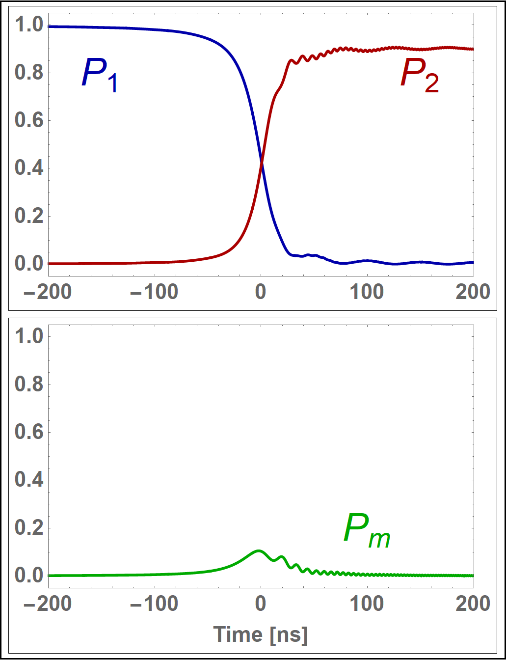}
 \caption{Coherent excitation transduction for the schematic in Fig. \ref{fig_QIS_design}(a). Temporal profiles of the populations $P_{1,2} = |c_{1,2}|^2$ of the cQED modes (top panel) and magnonic mode $P_m = |c_m|^2$ (bottom panel) in the case of linear sweep of the resonance magnon frequency $\omega_m(t) = \omega_0 + \rho t$. Adapted from Ref. \cite{TrevillianIntermag2020}.}
 \label{fig_OU_simulation}
\end{figure}

Fig. \ref{fig_OU_simulation} shows the results of numerical simulations of Eqs. (\ref{eq:nonstationary}) for the case of linear sweep of the magnon frequency ($\omega_m(t) = \omega_0 + \rho t$) with the rate $\rho/2\pi = 2.5$ MHz/ns. In the simulations, two cavities have a frequencies difference of $(\omega_2 - \omega_1)/2\pi = 40$~MHz, the decoherence rates of the two cavity modes and magnon mode are set as $\kappa_1/2\pi=\kappa_2/2\pi=0$ and $\kappa_m/2\pi = 1$~MHz, respectively, and the coupling rates are $g_{1m}/2\pi = g_{2m}/2\pi = 20$ MHz. Initially, at $t = -200$ ns, only the first cavity mode was populated ($c_1 = 1$, $c_{2,m} = 0$). The top panel in Fig. \ref{fig_OU_simulation} shows the time dependence of the populations $P_{1,2} = |c_{1,2}|^2$ of the cavity modes, while the bottom panel demonstrates the behavior of the magnon mode $P_m = |c_m|^2$. As one can see from Fig. \ref{fig_OU_simulation}, the sweep of the magnon frequency $\omega_m(t)$ across the resonance region $\omega_m \approx \omega_{1,2}$ results in almost complete coherent transfer of information state from the first microwave cavity to the second one. Note, that the simulations were done for rather conservative estimation of the coupling frequencies $g_{1,2}/2\pi = 20$ MHz, which are about one order of magnitude lower than experimentally observed coupling rates in hybrid systems of similar geometry. It also obvious from Fig. \ref{fig_OU_simulation} that the relative magnon population $P_m$ stays rather low during whole transfer process ($P_m < 0.2$), which means that this type of the coherent information transfer is almost insensitive to the decoherence processes in the magnonic subsystem. Finally, we note that the process shown in Fig. \ref{fig_OU_simulation} achieves coherent transfer of information between cavity modes of different frequencies. All this illustrates versatility and practical potential of coherent information manipulation in non-stationary hybrid magnon-cQED systems.

Figure \ref{fig_QIS_design}(b) illustrates another interesting possibility of quantum information manipulation in hybrid magnon-cQED systems, by constructing a magnonic hybrid splitter that can be used to create entangled magnon states at the two right ports when a single magnon excitation is input from one left port. The entanglement is achieved in the central part of the circuit, which represents a directional magnon coupler and operates similarly to a half-transparent mirror in optics. The entangled magnon state will be transferred to the qubits in the cQEDs via magnon-photon coupling, leading to entangled qubit states. The same circuit, when excited at both input ports with single-magnon excitations, can be used for generation of multi-magnon entangled states due to two-magnon interference effect (“magnon bunching”) in the directional magnon splitter. The two main advantages of using magnons compared with directly using microwave circuits are that 1) the magnon frequency can be tuned electronically in a wide range; 2) magnonic circuits can be made much smaller in dimension, in the scale of microns, compared with millimeters-cale microwave circuits.

\subsection{Magnonic logic}

\begin{figure}[htb]
 \centering
 \includegraphics[width=3.0 in]{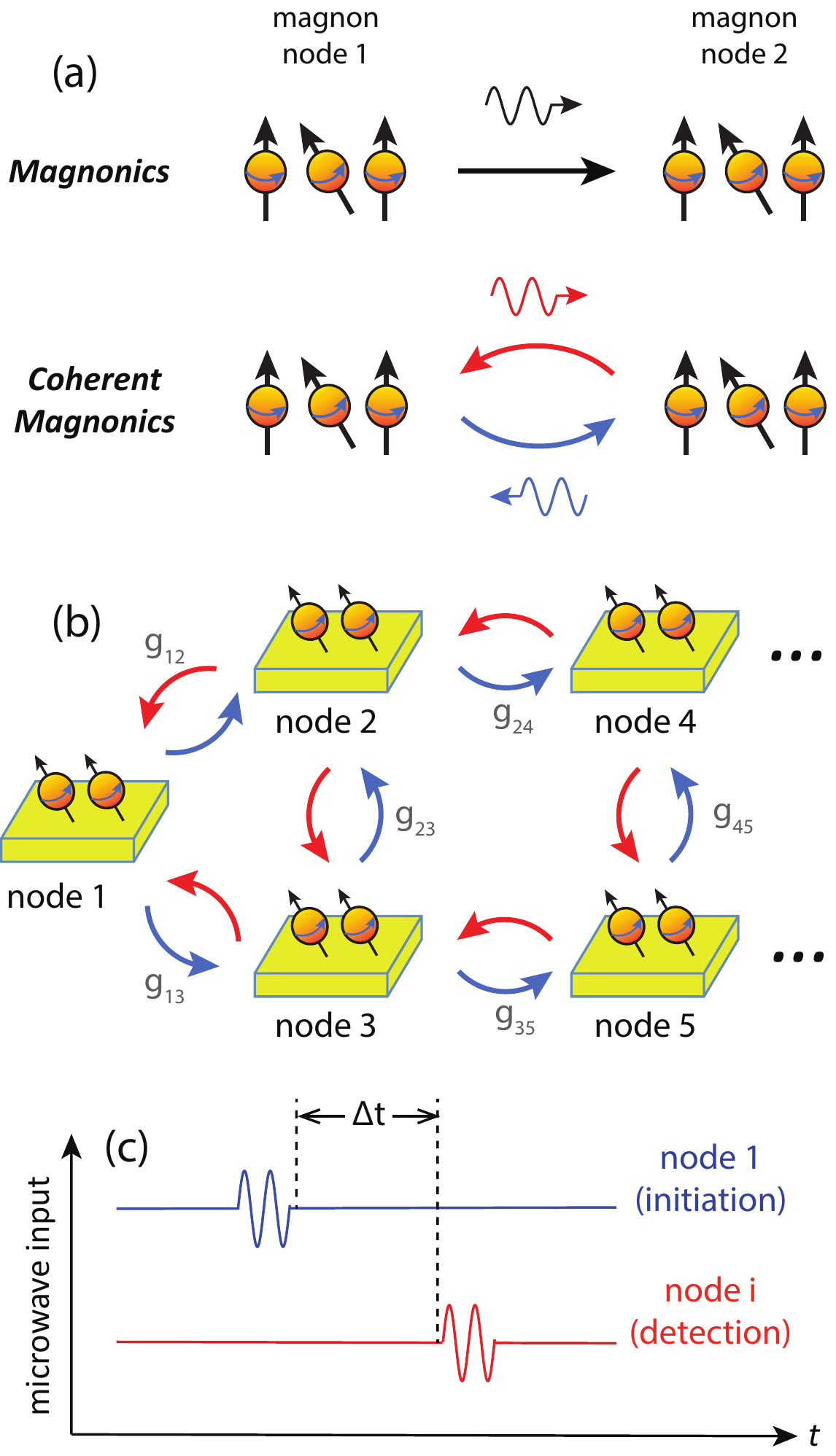}
 \caption{Proposed functionality in coherent magnonics with magnon hybrid system. (a) Comparison between conventional magnonics and coherent magnonics. (b) Network of magnon hybrid systems for coherent information processing. (c) Diagram of microwave pulse sequences for state initiations and detections.}
 \label{fig_coherent_magnonics}
\end{figure}

One major task of magnonics is to encode, transfer and process information with magnons \cite{KruglyakJPD2010,ChumakNPhys2015}, treating magnonic circuits as microwave circuits but making use of their tunability and nanoscale solid-state integration. With many device concepts based on magnon interference and phase control, coherent generations and manipulations of magnons are crucial for these operations. Magnon hybrid systems are meant for maintaining coherence during excitation transduction. Enhanced coupling into the strong-coupling regime will bring in new features such as Rabi-like oscillations \cite{ZhangPRL2014}. In particular, coherent remote information transduction mediated by magnon hybrid systems will greatly enhance the functionality of magnonics.

Let us compare conventional magnonics with \textit{coherent} magnonics based on magnon hybrid systems, as shown in Fig. \ref{fig_coherent_magnonics}(a). For conventional magnonics, the working mechanism takes a signal generation-detection process, i.e., magnons excited at one side, propagating and interfering with each other, and being received at the other side. For coherent magnonics, the working mechanism is manifested as a Rabi-like process: magnons that are excited from one node propagate and excite another node, then due to the strong coupling the excitation of the second node will generate a reverse flow of excitation back to the first node. This process will take place back and forth until the excitation flow completely decays or loses its coherence. By exchanging excitations, the two nodes will maintain coherence and act like entangled nodes. Although the number of magnon excitations are much greater than one, all the bosonic magnons are staying in the lowest energy state and cannot be distinguished, which is thus equivalent to entangled single magnon states between the two nodes.

Coherent interaction between the signal generator and detector will create new concepts for device implementation. In conventional magnonics, one typical device architecture is the spin wave logic gate \cite{SchneiderAPL2008}, in which binary states of magnon propagation are determined by the gate input. In coherent magnonics, one potential architecture is a network consisting of many magnon nodes that are coherently coupled [Fig. \ref{fig_coherent_magnonics}(b)]. Similar to quantum information, an initiation pulse will inject excitations into one magnonic nodes. Rabi-like oscillation will transfer the excitations to other nodes back and forth as a function of time. The excitation states of different nodes, including the amplitudes and phases, can be detected by spin rectification with a second detecting microwave pulse with designed time delay $\Delta t$ [Fig. \ref{fig_coherent_magnonics}(c)]. This is similar to the Ramsey interferometry \cite{RamseyPR1950,WilliamOliverAPRev2019} but the role of the second $\pi$ pulse is played by the detecting pulse. In quantum information, the calculation process is to take the coupling strengths among different nodes as input and measure the ground-state eigenfunction as output. For coherent magnonics, the coupling strength ($g_{ij}$) can be conveniently modified for input. The convergence to a specific ground state can be achieved by dissipative coupling (e.g., coherent spin pumping favors in-phase eigenmode and damps out-of-phase eigenmode \cite{LiPRL2020_YIGPy}). As an example, up to eight magnonic nodes have been demonstrated for YIG spheres placed inside a 3D cavity \cite{ZhangNComm2015}. There are also many ways in spintronics for rectification readout such as spin Hall effect, Rashba effect and anisotropic magnetoresistance. Furthermore, we note that Rabi-like oscillation, or nutation dynamics of magnons, has been realized \cite{CapuaNComm2017,LiPRX2019}, especially in a YIG single nanodisc with very large cone angle \cite{LiPRX2019}. This opens new opportunity for exploring coherent manipulation of magnonic states with large evolution bandwidth (nutation frequency) along with long coherence time, shown as 0.3~GHz for the former and 100~ns (0.01~GHz) for the latter in a 700-nm-diameter YIG nanodisc \cite{LiPRX2019}. The main challenge for coherent magnonics is the relatively short coherence time of magnons. However, although the damping of magnons are much larger than quantum qubits, the coupling strength can also be larger, resulting in a significant cooperativity for coherent information processing. Any realistic coherent magnonic system can be readily extended to the single quantum limit for coupling with qubits.

\section{Conclusion}

Coherent information processing with hybrid systems has become a new scientific challenge and inspired lots of interdisciplinary research activities especially for their application in quantum information. The introduction of magnons to the family has provided new freedom for implementing many dynamic phenomena that have been predicted in theory or realized in other dynamic systems, thanks to the convenience in tuning key properties such as frequency, damping and coupling. For magnetic systems, we have witnesses a tremendous success in turning great science into great applications, from the discovery of giant magnetoresistance (GMR) to the wide use of hard-disc drives. One important factor for this success is the convenience of integrating magnetic thin-film, heterostructure devices on circuits for scalable fabrications. Thus on-chip capability of magnon hybrid systems will be crucial for practical applications. Finally we anticipate an important role of magnon hybrid systems in quantum information and magnonics, namely, quantum magnonics and coherent magnonics.

\section{Acknowledgement}

The preparation of this perspective article was supported by the U.S. Department of Energy, Office of Science, Basic Energy Sciences, Materials Sciences and Engineering Division. W. Z. acknowledges support from AFOSR under grant no. FA9550-19-1-0254. A. H. acknowledges support from Quantum Materials for Energy Efficient Neuromorphic Computing, an Energy Frontier Research Center funded by the U.S. DOE, Office of Science.

\end{document}